\documentclass[12pt,a4paper]{jpconf}
\usepackage{graphicx}
\usepackage{sidecap}
\usepackage{multirow}

\newcommand{\SNeIa}{SNe~Ia}

\newcommand{\unitspace}{\ensuremath{\,}}
\newcommand{\usp}{\unitspace}
\newcommand{\numberspace}{\ensuremath{\;}}
\newcommand{\nsp}{\numberspace}
\newcommand{\unitstyle}[1]{\ensuremath{\mathrm{#1}}}
\newcommand{\power}[2]{\ensuremath{{#1}^{#2}}}

\newcommand{\cm}{\unitstyle{cm}}
\newcommand{\gram}{\unitstyle{g}}

\newcommand{\grampercc}{\gram\usp\power{\cm}{-3}} 

\newcommand{\yr}{\unitstyle{yr}}        

\newcommand{\pv}{\ensuremath{\phi}}

\newcommand{\Ni}[1]{{}^{#1}{\rm Ni}}
\newcommand{\mrow}[1]{\multirow{4}{*}{#1}}
\newcommand{\mcol}[1]{\multicolumn{2}{|r|}{#1}}

\newcommand{\araa}{Annu. Rev. Astron. Astrophys}
\newcommand{\aap}{Astronomy and Astrophysics}
\newcommand{\apj}{Astrophysical Journal}
\newcommand{\apjl}{Astrophysical Journal Letters}
\newcommand{\apjs}{Astrophysical Journal Supplement}
\newcommand{\prl}{Physical Review Letters}
\newcommand{\aj}{Astronomical Journal}
\newcommand{\mnras}{Monthly Notices of the Royal Astronomical Society}
\newcommand\nphysa{Nucl.~Phys.~A}
\newcommand\apss{Ap\&SS}

\begin{document}
\title{Evaluating Systematic Dependencies of Type Ia Supernovae}

\author{A.~C. Calder$^{1,2}$, B.~K.~Krueger$^1$, A.~P.~Jackson$^1$, 
D.~M.~Townsley$^3$, F.~X.~Timmes$^4$, E.~F.~Brown$^5$, and D.~A.~Chamulak$^6$}

\address{$^1$ Department of Physics and Astronomy, Stony Brook University, Stony Brook, NY 11794-3800}
\address{$^2$ New York Center for Computational Sciences, Stony Brook University, Stony Brook, NY, 11794-3800}
\address{$^3$ Department of Physics and Astronomy The University of Alabama, Tuscaloosa, AL, 35487-0324}
\address{$^4$ School of Earth and Space Exploration, Arizona State University, Tempe, AZ, 85287-1404}
\address{$^5$ Department of Physics and Astronomy, Michigan State University, East Lansing, MI 48824-2320}
\address{$^6$ Physics Division, Argonne National Laboratory, Argonne, IL 60439}

\ead{acalder@mail.astro.sunysb.edu}

\begin{abstract}
Type Ia supernovae are bright stellar explosions thought to occur when
a thermonuclear runaway consumes roughly a solar mass of degenerate
stellar material. These events produce and disseminate iron-peak elements, and
properties of their light curves allow for standardization and subsequent
use as cosmological distance indicators. The explosion mechanism of these
events remains, however, only partially understood.  Many models posit
the explosion beginning with a deflagration born near the center of a
white dwarf that has gained mass from a stellar companion. In order to
match observations, models of this single-degenerate scenario typically
invoke a subsequent transition of the (subsonic) deflagration to a
(supersonic) detonation that rapidly consumes the star. We present an
investigation into the systematics of thermonuclear supernovae assuming
this paradigm.  We utilize a statistical framework for a controlled study
of two-dimensional simulations of these events from randomized initial
conditions. We investigate the effect of the composition and thermal
history of the progenitor on the radioactive yield, and thus brightness,
of an event. Our results offer an explanation for some observed trends
of mean brightness with properties of the host galaxy.
\end{abstract}

\section{Introduction}
Type Ia supernovae (\SNeIa) are bright stellar explosions distinguished 
principally by a lack of hydrogen and strong silicon features in 
their spectra (for reviews, see \cite{Fili97,hillebrandt.niemeyer:type}).
Properties of the light curves of these events
allow their use as distance indicators at cosmological 
distances~\cite{phillips:absolute,reispreskirs+96,albrecht_2006_aa},
and these are at present the most powerful and best proved technique 
for studying dark energy~\cite{riess.filippenko.ea:observational,
perlmutter.aldering.ea:measurements,KolbReport-of-the-D,Hicketal09b,
Lampeitl10:SDSS}. Accordingly, there are many observational campaigns
underway striving to gather information about the systematics of these
events and to better measure the expansion history of the Universe
(see~\cite{Kirshner09} and references therein). These events are also
responsible for producing many of the heavy (iron-group) 
elements found in the galaxy and are therefore critical to
galactic chemical evolution~\cite{trucam71}.

Despite their widespread use as distance indicators and 
their importance to galactic chemical evolution, much of what is 
known about \SNeIa\ follows from empirical relationships, not a 
theoretical understanding of the explosion mechanism. 
Motivated principally by cosmological studies, observational 
campaigns are uncovering the rich phenomenology of these events at an
unprecedented rate and the future promises even more (the Dark Energy 
Survey, LSST, JDEM, see \cite{Howeetal09}). Accordingly, the 
goal of modeling \SNeIa\ is a theoretical understanding
of the observed  properties, particularly the intrinsic 
scatter of these events and the source of any systematic trends. 
Understanding and quantifying these are essential to
their effective use as distance indicators~\cite{albrecht_2006_aa,gehrels_2009_aa}.

A widely accepted view is that \SNeIa\  are the 
thermonuclear incineration of a white dwarf composed 
principally of carbon and oxygen that has gained mass from a 
companion star~\cite{hillebrandt.niemeyer:type}.
In this scenario, the white dwarf gains mass, compressing the core
until an explosion ensues.  Recent observational 
evidence, however, does suggest other progenitors such as the merging of
two white dwarf may explain many events \cite{scalzo:2010,Yuan:2010,gilfanov:2010}.  
Regardless of the exact mechanism, the ``upshot" is that these 
events synthesize $\sim 0.6 M_{\odot}$ of radioactive $^{56}$Ni, 
and the decay of this radioactive $^{56}$Ni,
not the explosion energy, powers the light curve. 
There is a
correlation, obeyed by the vast majority of \SNeIa, between the peak
brightness and the timescale over which the light curve decays from its
maximum.  This ``brighter is broader'' relation
\cite{phillips:absolute} explains the property of these events
that allows their use as distance indicators---by observing the decline from
maximum light, one can infer the peak brightness. The correlation 
is understood physically as stemming from having both the luminosity 
and opacity being set by the mass of $^{56}$Ni synthesized in the 
explosion~\cite{arnett:type,pinto.eastman:physics,Kasen2007On-the-Origin-o}. 

\subsection{Observational trends of \SNeIa}

Many contemporary observations address correlations between the event
and properties of the progenitor galaxy. Of particular interest are 
correlations between the brightness of an event and the isotopic 
composition of a galaxy and its age, measured by the intensity of
star formation. 

The proportion of material that has previously been processed in 
stars, i.e.\ elements other than hydrogen and helium, which are 
collectively referred to as ``metals'', is a measurable property 
of the galaxy. (The  relative abundance of these elements is referred 
to as the galaxy's ``metallicity.'') 
The presence of these elements in a progenitor white dwarf influences the 
outcome of the explosion by 
changing the path of nuclear burning, which influences the amount of 
$^{56}$Ni synthesized in an event. Because the decay of $^{56}$Ni 
powers the light curve, metallicity can directly influence the brightness of
an event.  Observational results to date are consistent with a shallow dependence
of brightness on metallicity, with dimmer events in metal-rich galaxies, but are 
unable demonstrate a conclusive trend~\cite{GallGarnetal05,gallagheretal+08,neilletal+09,howelletal+09}.
Determining the metallicity dependence is challenging because the
effect appears to be small, is difficult to measure, and 
there are systematic effects associated with the
mass-metallicity relationship within galaxies~\cite{gallagheretal+08}.
This effect is also difficult to decouple from the apparently
stronger effect of the age of the parent stellar population on the mean
brightness of SNe~Ia~\cite{gallagheretal+08,howelletal+09,Krueetal10}.

When galaxies form, they are rich in hydrogen gas and undergo a period
of intense star formation. Observations also target correlations between 
the brightness of an event and the age of a galaxy measured as the elapsed
time from the period of intense star formation. Some observations indicate
that the dependence of the SN Ia rate on delay time (elapsed time between 
star formation and the supernova event) is best fit by a bimodal distribution
with a prompt component less than 1~Gyr after star formation and a tardy
component several Gyr later~\cite{MannucciEtAl06,RaskinEtAl09}. 
Other studies only indicate a correlation between the delay time and brightness 
of \SNeIa\, with dimmer events occurring at longer delay 
times~\cite{GallagherEtAl08,howelletal+09,neilletal+09,BrandtEtAl10}.

\subsection{The deflagration to detonation transition model}

We explore the systematics of these events with models that
assume the explosion occurs in a carbon-oxygen white dwarf that has
gained mass from a companion.  Here we briefly describe this explosion
scenario and the physics involved.

The core of a white dwarf is dense enough that electrons are subject 
to quantum mechanical effects,
specifically the Pauli exclusion principle, that prevents the
electrons from occupying the same quantum states~\cite{dandw+60}.
The result is that electrons are forced into higher energy
states, and the reaction to this forcing acts as a pressure
that supports the star.  Matter in this high-density condition
is said to be ``degenerate'', and the case of interest, a
white dwarf that has gained mass from a companion, is known
as the ``single degenerate'' scenario.

If a white dwarf gains mass from a binary companion, the central 
density and temperature increase, and the star may eventually 
collapse to a neutron star. For a white dwarf composed 
principally of carbon and oxygen, before the collapse can 
occur the density and temperature reach values at which
carbon fusion begins and the star enters a period of simmering
that drives convection in the core producing a growing
convective zone~\cite{woosleyetal+04,zingale+09}.  After $\approx 10^{3}\nsp\yr$
the local temperature is hot enough that the burning timescale becomes 
shorter than a convective turnover time so that a local patch runs 
away and a subsonic flame is born, initiating the explosion~\cite{zingale+09}.

Early one-dimensional simulations of the single-degenerate case showed that 
the most successful scenario follows the initial deflagration (subsonic 
reaction front) with a (supersonic) detonation, i.e. a deflagration-detonation
transition~(DDT,~\cite{Khokhlov1991Delayed-detonat,HoefKhok96}).
Models with such a delayed detonation naturally account for some spectral 
features and the chemical stratification observed in the ejecta~\cite{kasen_ca}.
While one-dimensional models are able to reproduce observed features of the
light curve and spectra, much of the physics is missing.  
Of particular concern is the degree to which the white
dwarf expands during the deflagration phase of the explosion, which
multidimensional simulations show depends on the behavior of
fluid instabilities at the flame front. The degree of expansion during 
the deflagration phase is critical to the explosion because it determines the 
density at which the majority of the stellar material burns, which in turn 
controls the nucleosynthetic yield. Capturing the effects of fluid instabilities
is therefore essential to modeling this process and necessitates the
development of multidimensional models. By relaxing the symmetry constraints
on the model, buoyancy instabilities are naturally captured leading
to a strong dependence on the initial conditions of the deflagration.
Some cases with a DDT criterion based on previous one-dimensional
studies indicated the result is too little expansion of the star
prior to the detonation~\cite{NiemHillWoos96,
calder.ea:_offset_ignition_1,calder.ea:_offset_ignition_2,
Livne2005On-the-Sensitiv}. 
Multidimensional models, however, may reach the
expected amount of expansion prior to the DDT with the choice of
particular ignition conditions and thus retain the desirable features
from one-dimensional models~\cite{gamezo.khokhlov.ea:deflagrations,
PlewCaldLamb04,RoepGiesetal06,Jordan2008Three-Dimension}.
Our investigation centers on how properties of the progenitor white
dwarf that follow from properties of the host galaxy or its
evolutionary history influence this DDT scenario and hence the brightness
of an event measured by the $^{56}$Ni yield.

\section{The Systematics of \SNeIa\ in the DDT Scenario}

In Townsley et al.\ (2009)~\cite{townetal09} we investigated the 
direct effect of reprocessed stellar material (metals) in the host
galaxy via the initial neutron excess of the progenitor white dwarf. 
Because of weak interactions, metals produced by nuclear burning are 
more neutron rich than hydrogen and helium, and, accordingly, the 
neutron excess of these elements is thought to drive the explosion 
yield toward stable iron-group elements. Thus, there is relatively less 
radioactive $^{56}$Ni in the NSE mix, which results in a dimmer 
event~\cite{timmes.brown.ea:variations}. 
We investigated this effect by introducing $^{22}$Ne into the
progenitor white dwarf as a proxy for (neutron-rich) metals. 
The presence of $^{22}$Ne influences the progenitor structure, the energy
release of the burn, and the flame speed.  The study was designed to measure 
how the $^{22}$Ne content influences the competition between rising plumes 
and the expansion of the star, which determines the yield.
We performed a suite of 20 DDT simulations varying only the
initial $^{22}$Ne in a progenitor model, and found a
negligible effect on the pre-detonation expansion of the star
and thus the yield of NSE material.  The neutron excess sets the 
amount of material in NSE that favors stable iron-group elements over 
radioactive $^{56}$Ni. Our results were consistent with earlier 
work calculating the direct modification of $^{56}$Ni mass from
initial neutron excess~\cite{timmes.brown.ea:variations}.

In Jackson et al.\ (2010)~\cite{jacketal} we expanded the
Townsley et al.\ study to include the indirect effect of metallicity in
the form of the $^{22}$Ne mass fraction through its influence
on the density at which the DDT takes place. We simulated 30 realizations 
each at 5 transition densities between $1-3\times10^7$~g~cm$^{-3}$ for a 
total of 150 simulations.  We found a quadratic dependence of the NSE 
yield on the log of the transition density, which is determined by the 
competition between rising unstable plumes and stellar expansion. By 
then considering the effect of metallicity on the transition density, we 
found the NSE yield decreases slightly with metallicity, but that the ratio 
of the $^{56}$Ni yield to the overall NSE yield does not change as 
significantly. Observations testing the dependence of the yield on metallicity 
remain somewhat ambiguous, but the dependence we found is comparable to that 
inferred from~\cite{bravo10}.  We also found that the scatter in the results 
increases with decreasing transition density, and we attribute this increase in
scatter to the nonlinear behavior of the unstable rising plumes.

In Krueger et al.\ (2010)~\cite{kruegetal} we investigated
the effect of central density on the explosion yield.
We found that the overall production of NSE material did not
change, but there was a definite trend of decreasing $^{56}$Ni 
production with increasing progenitor central density.
We attribute this result to higher rates of weak interactions
(electron captures) that produce a higher proportion
of neutronized material at higher density.  More neutronization
means less symmetric nuclei like $^{56}$Ni, and, accordingly, 
a dimmer event. This result may explain
the observed decrease in \SNeIa\ brightness with increasing delay
time. The central density of the progenitor is determined by
its evolution, including the transfer of mass from the companion. 
If there is a long period of cooling before the onset of mass transfer, 
the central density of the progenitor will be higher when the core 
reaches the carbon ignition temperature~\cite{Lesaffre2006The-C-flash-and}, 
thereby producing less $^{56}$Ni and thus a dimmer event.  In addition, we found
considerable variation in the trends from some realizations, stressing the
importance of statistical studies.

Our approach in these investigations has been to isolate 
facets of the problem expected to follow from properties of the host
galaxy or evolution history of the progenitor white dwarf and
perform a controlled statistical analysis of a ensemble of 
multidimensional simulations (described below).  Not surprisingly, 
many other possible systematic effects exist (outlined in \cite{townetal09})
that were held fixed in each study. The ultimate goal of
our research into thermonuclear supernovae is to consider the 
interdependence of all of these effects in the construction of the 
full theoretical picture.

\section{Methodology}

Our methodology for the theoretical study of thermonuclear
supernovae consists of four principal parts. First is
the ability to construct parameterized, hydrostatic initial white
dwarf progenitors that can freely change thermal and compositional 
structure to match features from the literature about 
progenitor models~\cite{jacketal}. Second is a model flame and energetics
scheme with which to track both (subsonic) deflagrations and (supersonic) 
detonations as well as the evolution of dynamic ash in NSE.
This flame/energetics scheme is implemented 
in the Flash hydrodynamics 
code~\cite{Fryxetal00,calder.curtis.ea:high-performance,calder.fryxell.ea:on}.
Third is utilization of a scheme to post-process the density and temperature 
histories of Lagrangian tracer particles with a detailed nuclear
network in order to calculate detailed nucleosynthetic 
yields~\cite{brown.calder.ea:type,townetal10}. Fourth, we developed a 
statistical framework with which to perform ensembles of simulations for
well-controlled studies of systematic effects. 
Below we highlight the flame model and the statistical 
framework. Complete details of the methodology can be found in
previously published results~\cite{brown.calder.ea:type,Caldetal07,townsley.calder.ea:flame,townetal09,townetal10}. 

\subsection{Flame model}
The great disparity between the length scale of a white
dwarf ($\sim 10^9\ensuremath{\;}{\ensuremath{\mathrm{cm}}}$) and the width
of laminar nuclear flame ($< 1\ensuremath{\;}{\ensuremath{\mathrm{cm}}}$) 
necessitates the use of a model flame in simulations of thermonuclear
supernovae. Even simulations with adaptive mesh refinement cannot 
resolve the actual diffusive flame front in a simulation of the event.
The model we use propagates an artificially
broadened flame front with an advection-diffusion-reaction (ADR)
scheme~\cite{Khok95,VladWeirRyzh06} that has been demonstrated to be
acoustically quiet and produce a unique flame speed~\cite{townsley.calder.ea:flame}. 
This scheme evolves
a reaction progress variable $\phi$, where $\phi=0$ indicates unburned fuel
and $\phi=1$ indicates burned ash, with the
advection-reaction-diffusion equation
\begin{equation}
  \label{eq:ard}
  \partial_t \pv + \vec{u}\cdot\nabla \pv = \kappa \nabla^2 \pv + 
\frac{1}{\tau} R\left(\phi\right).
\end{equation}
Here $\kappa$ is a constant and $R(\phi)$ a non-dimensional function, and both 
are tuned to propagate the reaction front at the physical speed of the
real flame~\cite{timmes92,ChamBrowTimm07} and to be just wide
enough to be resolved in our simulation.  We use a modified version of the
KPP reaction rate discussed by~\cite{VladWeirRyzh06}, in which
$R\propto(\phi-\epsilon)(1-\phi+\epsilon)$, where $\epsilon \simeq 10^{-3}$.

In simulating the deflagration phase, the flame front separates expanded
burned material (the hot ash) from denser unburned stellar material (cold
fuel). The expansion and buoyancy of the burned material forces the interface
upward into the denser fuel, and the configuration is susceptible to 
the Rayleigh-Taylor fluid instability~\cite{taylor+50,chandra+81}. It is necessary to enhance the
flame speed in order to prevent turbulence generated by the Raleigh-Taylor
instability from destroying the artificially broadened flame front.
In the simulations discussed here, the enhancement is accomplished by the method 
suggested by Khokhlov~\cite{Khok95} in which we prevent the flame front speed, $s$, 
from falling below a threshold that is scaled with the strength of 
the Raleigh-Taylor instability on the scale of
the flame front ($s\propto \sqrt{g\ell}$, where $g$ is the local gravity and
and $\ell$ is the width of the artificial flame, which is a few times the
grid resolution).  This scaling of the flame speed prevents the Rayleigh-Taylor 
instability from effectively 
pulling the flame apart and also mimics what is a real enhancement of the burning 
area that is occurring due to structure in the flame surface on unresolved scales.  
It is expected that, for much of the white dwarf deflagration, the flame is
``self-regulated'', in which the small scale structure of the flame surface
is always sufficient to keep up with the large-scale buoyancy-driven motions
of the burned material.  Thus the actual burning rate is determined by this
action, which is resolved in our simulations. 

\begin{figure}[htbp]
\centering{\includegraphics[width=3in]{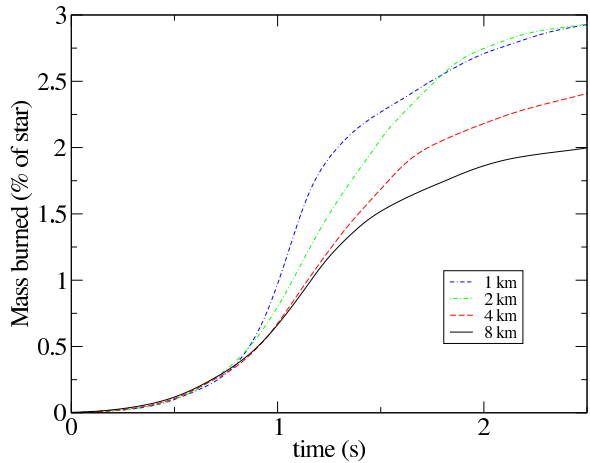}}\hfill\parbox[b]{3.1in}{
  \caption{
    \label{fig:convergence}
    Fraction of star burned during the deflagration phase of a thermonuclear
    supernova for four two-dimensional simulations at varying resolutions. The initial 
    conditions consisted of a 16 km ignition point started at a radius of 40 km from 
    the center of the star. The result of this off-set ignition point is a single
    rising plume that may subsequently trigger a detonation~\cite{PlewCaldLamb04}.
    Shown are effective resolutions of 8, 4, 2, and 1 km, and the results
    demonstrate reasonable convergence, especially during the early phases. 
    Adapted from Townsley et al.\ (2009)~\cite{townsley.calder.ea:flame}.
  }}
\end{figure}

As seen in Figure~\ref{fig:convergence}, \cite{townsley.calder.ea:flame} and by
\cite{gamezo.khokhlov.ea:thermonuclear}, these techniques demonstrate a
suitable level of convergence for studies such as ours.  This technique
does makes it necessary to explicitly drop $s$ to zero below a density of
$10^7$ g cm$^{-3}$, approximately where the real flame will be extinguished.
This prescription captures
some effects of the Rayleigh-Taylor instability and maintains the integrity
of the thickened flame, but it does not completely describe the flame-turbulence
interaction. Also, we neglect any enhancement from background turbulence from
convection prior to the birth of the flame. Future work will include 
physically-motivated models for these effects~\cite{jackinprep}.

The model also includes the nuclear energy release occurring at the flame front and
in the dynamic ash in NSE. We performed a detailed study of the nuclear processes
occurring in a flame in the interior of a white dwarf and developed an efficient and
accurate method for incorporating the results into numerical simulations~\cite{Caldetal07}.
Tracking even tens of nuclear species is computationally prohibitive, and many 
more than this are required to accurately calculate the physics such as electron 
captures rates that are essential to studying rates of neutronization.  We instead 
reproduce the energy release of the nuclear reactions with a highly abstracted model 
based on tabulation of properties of the burned material calculated in
our study of the relevant nuclear processes. 

The nuclear processing can be well approximated as a three stage process:
initially carbon is consumed, followed by oxygen, which creates a mixture of 
silicon-group and light elements that is in nuclear statistical quasi-equilibrium
(NSQE), finally the silicon-group nuclei are converted to iron-group, reaching full NSE.  
In both of these equilibrium states, the capture and creation of light elements 
(via photodisintegration) is balanced, so that energy release can continue by 
changing the relative abundance of light (low nuclear binding energy) and heavy 
(high nuclear binding energy) nuclides, an action that releases energy as buoyant burned 
material rises and expands.
We track each of these stages with separate progress variables and separate
relaxation times derived from full nuclear network calculations
\cite{Caldetal07}.  We define three progress variables
representing consumption of carbon $\phi_{\rm C}$, consumption of oxygen to
material in NSQE, $\phi_{\rm NSQE}$, and conversion of silicon to iron group nuclides
to form true NSE material, $\phi_{\rm NSE}$.  The physical state of the fluid is
tracked with the electron number per baryon, $Y_e$, the number of nuclei
per baryon, $Y_{\rm i}$, and the average binding energy per baryon, $\bar q$,
the minimum properties necessary to hydrodynamically evolve the fluid.
Carbon consumption is coupled directly to the flame progress variable,
$\phi_{\rm C}\equiv\phi$, from eq.\ (\ref{eq:ard}) above, and the later flame
stages follow using simple relationships from more detailed calculations.

Burning and evolution of post-flame material change the nuclear binding
energy, and we use the binding energy of magnesium to approximate
the intermediate burning products of carbon. The method is finite
differenced in such a way to ensure explicit conservation of energy.
Weak processes (e.g.\  electron capture) are included in the calculation 
of the energy input rate, as are neutrino losses, which are calculated 
by convolving the NSE distribution with the weak interaction cross sections.
Tracking the conversion of silicon-group nuclides to the iron-group is important 
for studying the effects of electron capture because the thresholds are lower for
the iron-group nuclides.  Both the NSE state and the electron capture rates were
calculated with a set of 443 nuclides including all which have weak interaction 
cross sections given by \cite{langanke.martinez-pinedo:weak}.  This treatment of 
electron capture in the energy release is the most realistic currently in use for 
thermonuclear supernovae.  Electron capture feeds back on the hydrodynamics in 
three ways: the NSE can shift to more tightly bound elements as the electron
fraction, $Y_e$, changes, releasing some energy and changing the local
temperature; also the reduction in $Y_e$ lowers the Fermi energy, reducing
the primary pressure support of this highly degenerate material and having an
impact on the buoyancy of the neutronized material; finally neutrinos are
emitted (since the star is transparent to them) so that some energy is lost
from the system.

In addition to all of these effects during the deflagration phase
of \SNeIa, the progress-variable-based method has been extended to
model the gross features of detonations~\cite{Meaketal09,townetal10}. 
Instead of coupling the first burning stage, $\phi_{\rm C}$,
representing carbon consumption, to the ADR flame
front, we instead can use the actual temperature-dependent rate for carbon
burning, or possibly a more appropriate effective rate.  Doing so allows  shock
propagation to trigger burning and therefore create a propagating detonation.
This method has been used successfully by \cite{GameKhokOran05}
in modern studies of the DDT, and our
multistage burning model shares many features with theirs (see also
\cite{khokhlov91+dd} and \cite{khokhlov+00}).  We treat the later
burning stages very similarly, though we have taken slightly more care to
track the intermediate stages and have nearly eliminated acoustic noise when
coupling energy release to the flame. 

\begin{figure}[htbp]
\centering{\includegraphics[height=71mm]{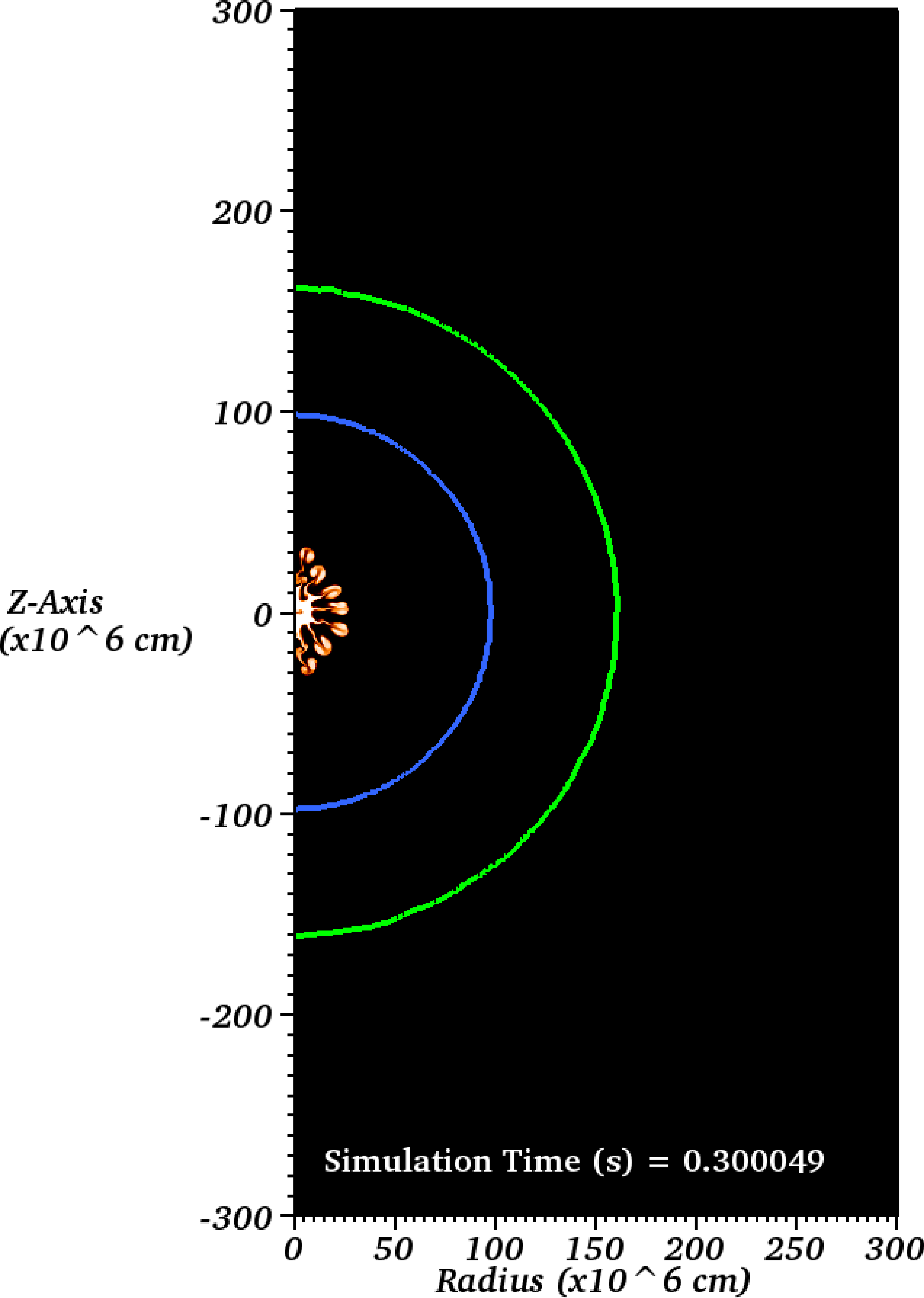}
\includegraphics[height=71mm]{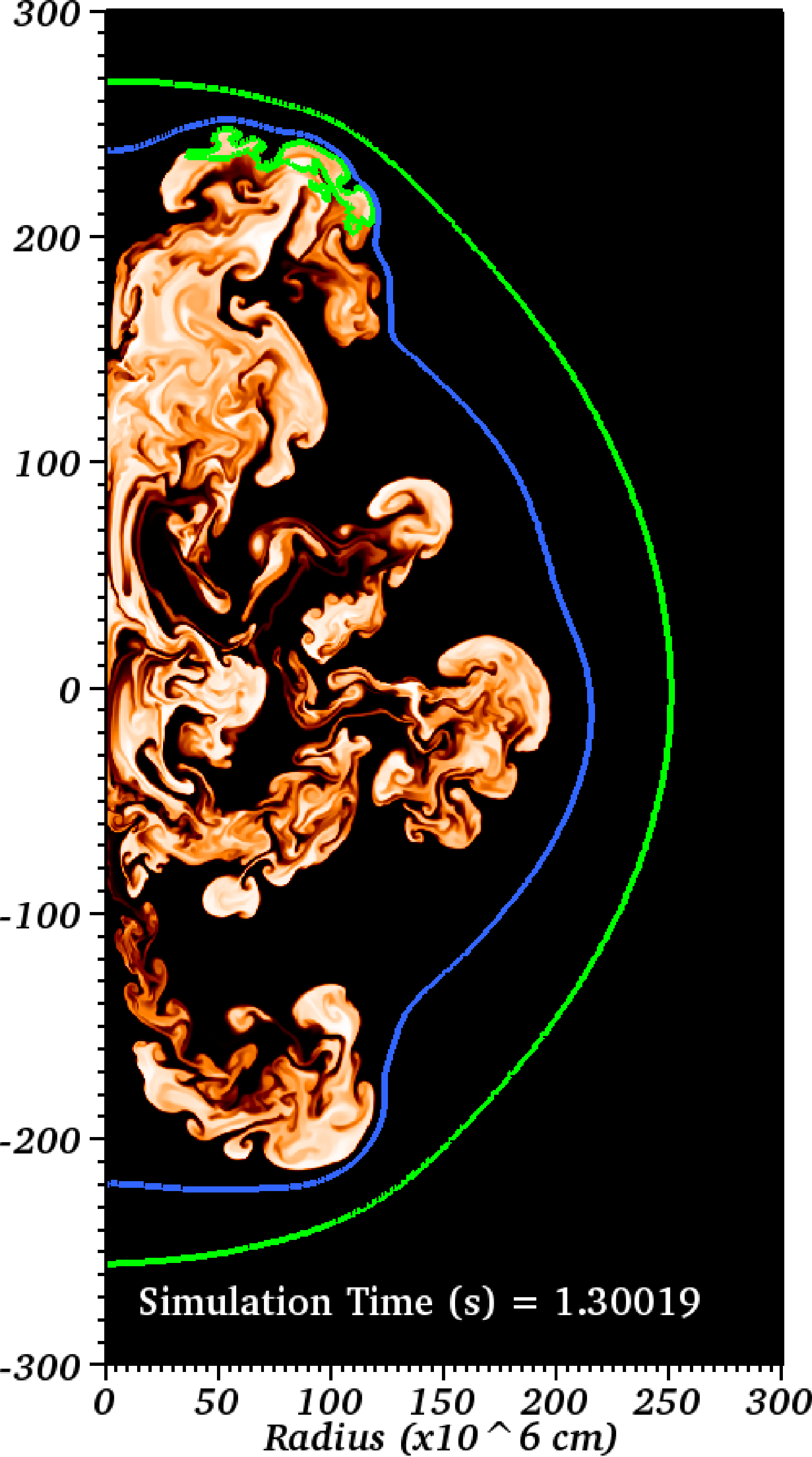}
\includegraphics[height=71mm]{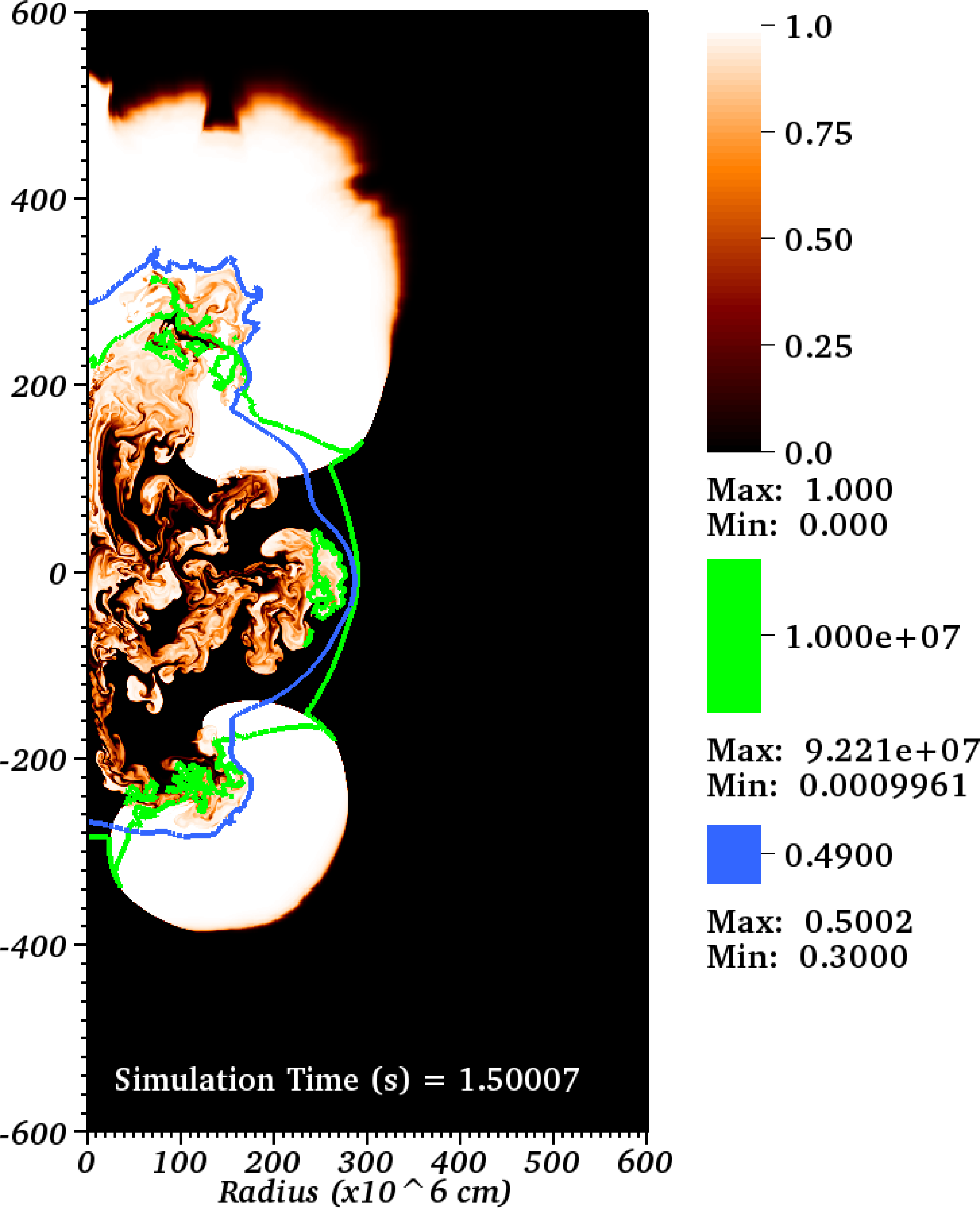}}
\caption{Images from a two-dimensional SNe Ia simulation from a neutronized-core progenitor.
The left panel shows the development of fluid instabilities during the early deflagration
phase, the center panel shows the configuration just prior to the first detonation,
and the right panel shows the configuration with two distinct detonations
consuming the star. Shown in color scale is the carbon-burning reaction progress variable which evolves
from 0 to 1 and contours of $\rho = 10^7 \grampercc$ (green) and where the initial
$X_{^{12}C} = 0.49$ (blue).  The latter, initially inner, contour indicates the separation
between the convective, low C-abundance, neutronized core and the higher C surface layers.
Note that the scale on the right panel is twice that of the first two in order to
accommodate the expansion of the star.}
\label{f:ddt}
\end{figure}

Results from a two-dimensional DDT simulation utilizing
these capabilities are shown in Figure~\ref{f:ddt}.
The three panels illustrate the early deflagration phase, the configuration 
just prior to ignition of the first detonation, and the progress of the detonation.
The blue contour marks the division between the pre-explosion convective core,
which has C/O/$^{22}$Ne composition of 30/66/4\% by mass, and an
outer region with a composition of 50/48/2\%.
The contrast in C/O abundance reflects the stellar evolution of the progenitor,
during which the central regions are produced by convective core helium burning,
while the outer layers are produced by shell burning during the asymptotic giant
branch phase~\cite{Straetal03}. The contrast in $^{22}$Ne is due to production during the
simmering phase. 

\subsection{Statistical Framework}

We also developed a theoretical framework for the study of systematic effects 
in \SNeIa\ that will utilize two- and three-dimensional simulations in the DDT 
paradigm~\cite{townetal09}.
This framework allows the evaluation of the average dependence of the properties 
of \SNeIa\ on underlying parameters,
such as composition, by constructing a theoretical sample based on a
probabilistic initial ignition condition.  Such sample-averaged dependencies are
important for understanding how SN~Ia models may explain features of the
observed sample, particularly samples generated by large dark energy
surveys utilizing \SNeIa\ as distance indicators.

The theoretical sample is constructed to represent statistical properties of
the observed sample of \SNeIa\ such as the mean inferred $\Ni{56}$
yield and variance. Within the DDT paradigm, the variance in $\Ni{56}$
yields can be explained by the development of fluid instabilities during
the deflagration phase of the explosion. By choice of the initial
configuration of the flame, we may influence the growth of these fluid
instabilities resulting in varying amounts of $\Ni{56}$ synthesized
during the explosion. \cite{townetal09} found that perturbing a
spherical flame surface with radius ($r_0 = 150$ km) with spherical
harmonic modes ($Y_l^m$)
between $12 \leq l \leq 16$ with random amplitudes ($A$) normally
distributed between $0-15$ km and, for 3D, random phases ($\delta$)
uniformly distributed between $\mbox{-}\pi-\pi$ best characterized the
mean inferred $\Ni{56}$ yield and sample variance from observations:
\begin{eqnarray}
\label{eq:init_surf}
r(\theta) &=& r_0 + \sum_{l=l_{\rm min}}^{l_{\rm max}} A_l Y_l(\theta) \\
r(\theta,\phi) &=& r_0 + \sum_{l=l_{\rm min}}^{l_{\rm max}} \sum_{m=-l}^l 
\frac{A_l^m e^{i\delta_l^m}}{\sqrt{2l+1}} Y_l^m(\theta,\phi)
{\rm .}
\end{eqnarray}
With a suitable random-number generator, a
sample population of progenitor WDs is constructed by defining the initial
flame surface for a particular progenitor.

\section{Statistical properties of the simulations}

For this contribution, we focus on the ensemble of simulations from
the central density study. In this study, the measure of the explosion
is the yield of radioactive $^{56}$Ni, the decay of which powers the 
light curve.  In comparing a simulated explosion to astronomical observations, 
the results from simulations must be considered at frequencies of light 
corresponding to the bands in which the observations are made. We applied a 
simple relationship between the mass of $^{56}$Ni synthesized in the
explosion and the peak brightness in the V band. Thus the assumption for 
these models that the amount of $^{56}$Ni synthesized in the explosion
directly corresponds to the peak V-band 
brightness~\cite{howelletal+09,goldhaberetal01,phillips.lira.ea:reddening-free}.

In the study, we found considerable variation in the amount of synthesized
$^{56}$Ni that we attribute to differences in the evolution during the
deflagration phase of the explosion. As described above, the initial
conditions for a realization were established stochastically. The deeply
non-linear behavior of the subsequent evolution the buoyant rising plumes
results in considerable variation in the duration of the deflagration
phase, which is set by the time required for the first plume to rise
to the DDT density. During the deflagration, the star reacts to the
subsonic burning and expands.  When a plume reaches the DDT threshold,
the subsequent detonation very rapidly incinerates the expanded star
(see Figure~\ref{f:ddt} for an illustration). $^{56}$Ni is synthesized
when stellar material burns at relatively high densities, so the
amount of expansion determines the mass of material that will burn to
$^{56}$Ni. Thus the evolution during the deflagration, particularly its
duration, strongly influences the $^{56}$Ni yield. From this ``noisy''
background we were able to find a trend of decreased $^{56}$Ni yield with
increasing central density.
\begin{figure}[htbp]
\centering{\includegraphics[width=3.4in,angle=270]{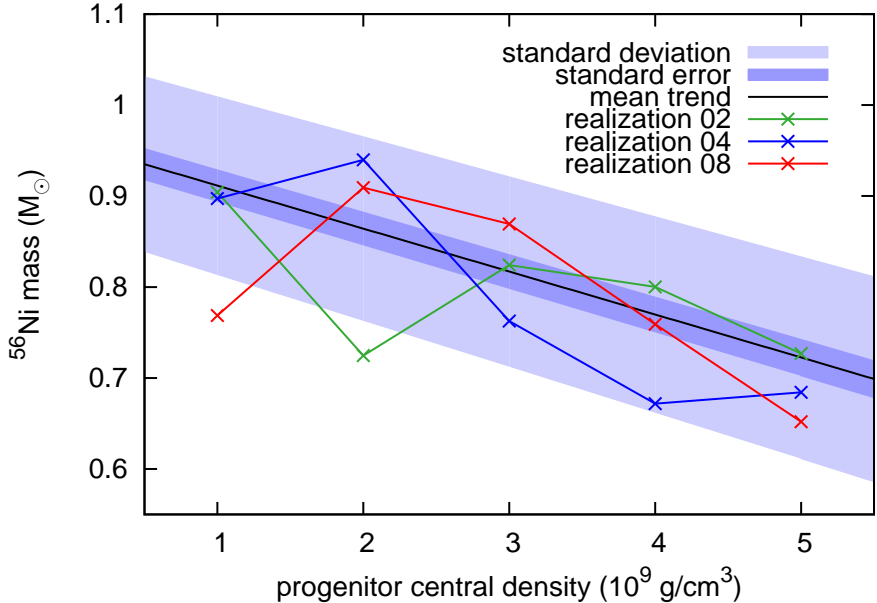}
  \caption{
    \label{fig:scatter}
    Relationship between mass of $^{56}$Ni produced in a simulated event and
    central density of the progenitor white dwarf when the deflagration is
    ignited.  The light purple is the standard deviation, the dark purple is
    the standard error of the mean, and the black line is the mean trend.  Also
    shown are the trends for three different initial configurations as the
    central density varies.  These three configurations do not match the
    statistical trend, nor are they monotonic; from this we conclude that a
    statistical study is important when considering a highly nonlinear problem
    such as a SN~Ia.  Adapted from Krueger et al.\ (2010)~\cite{kruegetal}.
  }}
\end{figure}

Figure~\ref{fig:scatter} plots the mass of $^{56}$Ni as a function of central 
density, illustrating the trend and scatter in our ensemble of simulations. 
Shown are the average among realizations,
the standard deviation, and the standard error. Note that the standard
deviation of our sample is about as large as the limits of the trend.
Realizations 2, 4, and 8 in the figure exemplify the
variation with progenitor central density, specifically the
non-monotonicity of the individual trends. While these three
realizations show a decreasing standard deviation with increasing
central density, when considering the entire sample population, the
standard deviation remains approximately the same as a function of
central density. Table~\ref{tab:fit} provides the minima, mean, maxima, and
standard deviations for $\Ni{56}$ and NSE masses at each central density.

\begin{SCfigure}
\centering
\includegraphics[width=2.7in,angle=270]{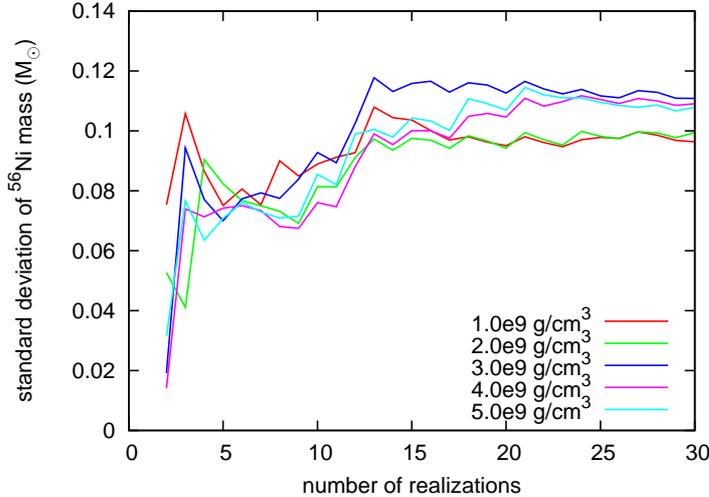}
\caption{\label{fig:stddev}
      Convergence of the standard deviation of the $^{56}$Ni mass as the data
      sample size increases for each of the five progenitor central densities.
      Each ``realization'' constitutes a unique initial condition
      configuration.  Around 15-20 realizations achieve approximately the final
      standard deviation.  The convergence of the standard deviation suggests
      that a sufficient number of data points have been included to compute a
      statistically correct mean.
  }
\end{SCfigure}

In order to obtain a statistically meaningful trend, we must first
demonstrate that our sample size characterizes the properties of the 
population. In Figure~\ref{fig:stddev}, we show that the standard
deviation of the $\Ni{56}$ mass converges for all progenitor central
densities with 15--20 realizations. Second, we must perform a number of
realizations (take enough samples) such that the standard error of the
mean is small enough to produce a statistically meaningful average
trend. In the case that no trend exists, the number or realizations
provides a limit to the magnitude of the mean trend. The standard
error of the mean is computed to be the standard deviation divided by
the square root of the number of samples.

We analyzed the trend of $\Ni{56}$ mass with progenitor central density
with 30 realizations. While not shown graphically, the NSE mass does not
vary significantly with central density. We evaluate an upper limit to
the trend to be a decrease of 0.006 solar masses of NSE per 
$10^9$~g~cm$^{-3}$ of central density. That result translates to less
than a 0.6\% change in the NSE yield per $10^9$~g~cm$^{-3}$.
From Figures~\ref{fig:scatter} and~\ref{fig:stddev}, we conclude that
a statistical approach is necessary to evaluate systematic trends in
highly nonlinear problems, such as \SNeIa. 
\begin{table}
\centering
\caption{\label{tab:fit}Minima, means ($\mu$), maxima, and standard deviations
($\sigma$) of $\Ni{56}$ and NSE masses from central density study.  All masses
are in units of $M_\odot$.}
\begin{tabular}{|l r|r r r r r|}
\hline \hline
\mcol{$\rho_c$ (g/cm$^3$)} &1.000E+9 &2.000E+9 &3.000E+9 &4.000E+9 &5.000E+9 \\
\hline
\mrow{$\Ni{56}$} &min      &6.920E-1 &6.905E-1 &5.657E-1 &5.828E-1 &5.304E-1 \\
                 &$\mu$    &9.105E-1 &8.723E-1 &8.035E-1 &7.713E-1 &7.255E-1 \\
                 &max      &1.052E+0 &1.034E+0 &9.813E-1 &9.640E-1 &9.288E-1 \\
                 &$\sigma$ &9.638E-2 &9.945E-2 &1.109E-1 &1.091E-1 &1.078E-1 \\
\hline
\mrow{NSE}       &min      &7.969E-1 &8.565E-1 &7.783E-1 &8.450E-1 &8.197E-1 \\
                 &$\mu$    &1.032E+0 &1.050E+0 &1.028E+0 &1.038E+0 &1.026E+0 \\
                 &max      &1.187E+0 &1.225E+0 &1.216E+0 &1.228E+0 &1.231E+0 \\
                 &$\sigma$ &1.039E-1 &1.044E-1 &1.149E-1 &1.101E-1 &1.082E-1 \\
\hline
\end{tabular}
\end{table}

\section{Conclusions}

Many studies have shown that small changes to the chosen initial
conditions of a \SNeIa\ model result in large changes to the outcome of
the explosion due to the non-linear effect of fluid instabilities. Our
statistical framework relies on this property of multi-dimensional \SNeIa\
simulations to
produce a sample population representative of the observed population.
However, for the first time, we have demonstrated that these small
changes to the initial conditions can also influence systematic trends
with properties of the progenitor WD such as central density. The effect
of varying central density within a single realization does not
characterize the entire sample population! In fact, each
realization's outcome varies in a different way with changing
central density as shown in Figure~\ref{fig:scatter}. It is only when all
realizations in the sample population are considered that a meaningful and
statistically significant trend emerges.

By varying the central density in a progenitor WD model, we also vary the
density structure. The speed of the laminar flame depends on the density
of the fuel it consumes. In hindsight, it is not that surprising that
varying the fuel density of the initial flame results in non-linear behavior
of the resulting supernovae similar to the effect of varying the initial
position of the flame surface. Therefore, we must use a statistical
ensemble of simulations to evaluate systematic trends in multi-dimensional
\SNeIa\ that are subject to fluid instabilities.

In future studies we plan to explore systematic effects due to varying the
core carbon-to-oxygen (C/O) ratio. The core C/O ratio is thought to vary
with zero-age main sequence mass of the progenitor WD, metallicity,
and the mass of the companion star~\cite{umeda.nomoto.ea:evolution}.
Because the laminar flame speed also depends on the carbon abundance in
the fuel, a statistical ensemble of simulations will be necessary to
evaluate this effect. It is not clear whether the results
of~\cite{RoepGiesetal06} are characteristic of the observed sample of
\SNeIa\ because a statistical approach was not employed.

\ack

This work was supported by the Department of Energy through grants
DE-FG02-07ER41516, DE-FG02-08ER41570, and DE-FG02-08ER41565, and by NASA
through grant NNX09AD19G.  ACC also acknowledges support from the
Department of Energy under grant DE-FG02-87ER40317. DMT received support
from the Bart J. Bok fellowship at the University of Arizona for part of
this work. The authors acknowledge the hospitality of the Kavli Institute 
for Theoretical Physics, which is supported by the NSF under grant 
PHY05-51164, during the programs ``Accretion and Explosion: the Astrophysics 
of Degenerate Stars'' and ``Stellar Death and Supernovae.''  The software 
used in this work was in part developed by the DOE-supported ASC/Alliances 
Center for Astrophysical Thermonuclear Flashes at the University of Chicago. 
This work was also supported in part by the US Department of Energy, 
Office of Nuclear Physics, under contract DE-AC02-06CH11357 and 
utilized resources at the New York Center for Computational Sciences 
at Stony Brook University/Brookhaven National Laboratory which is 
supported by the U.S.  Department of Energy under Contract No. 
DE-AC02-98CH10886 and by the State of New York.

\section*{References}

\end{document}